\documentclass[prd,aps,
nofootinbib,%
twocolumn,
superscriptaddress
]{revtex4-2}

\usepackage{graphicx}
\usepackage{amsmath}
\usepackage{amssymb}
\usepackage{hyperref} 

\usepackage{color}

\usepackage{colortbl}

\newcommand{\be}{\begin{equation}}
\newcommand{\ee}{\end{equation}}
\newcommand{\bea}{\begin{eqnarray}}
\newcommand{\eea}{\end{eqnarray}}

\newcommand{\C}{\mathcal{C}}
\newcommand{\F}{\mathcal{F}}
\newcommand{\J}{\mathcal{J}}

\newcommand{\beq}{\begin{equation}}
\newcommand{\eeq}{\end{equation}}

\newcommand{\Q}{\mathcal{Q}}
\newcommand{\M}{\mathcal{M}}

\def\HS{\cite{Hoferichter:2020lap}}

\begin{document}

\title{Longitudinal short-distance constraints on
hadronic light-by-light scattering\\ and tensor meson contributions to the muon $g-2$
}

\author{Jonas Mager}
\affiliation{Institut f\"ur Theoretische Physik, Technische Universit\"at Wien,
        Wiedner Hauptstrasse 8-10, A-1040 Vienna, Austria}
\author{Luigi Cappiello}
\affiliation{Dipartimento di Fisica "Ettore Pancini", Universit\`a di Napoli "Federico II"} 
\affiliation{
INFN-Sezione di Napoli, Via Cintia, I-80126 Napoli, Italy}
\author{Josef Leutgeb}
\author{Anton Rebhan}
\affiliation{Institut f\"ur Theoretische Physik, Technische Universit\"at Wien,
        Wiedner Hauptstrasse 8-10, A-1040 Vienna, Austria}

\date{\today}

\begin{abstract}
Short-distance constraints from the operator product expansion in QCD play an
important role in the evaluation of the hadronic light-by-light scattering
contribution to the anomalous magnetic moment of the muon. Holographic QCD has
been shown to naturally incorporate the Melnikov-Vainshtein constraint on
the longitudinal amplitude following from the triangle anomaly
in the asymmetric limit, where one photon virtuality
remains small compared to the others. This is saturated by an
infinite tower of axial vector mesons, and their numerical contribution
to the muon $g-2$ in AdS/QCD models agrees rather well with
a recent dispersive analysis. However,
in AdS/QCD,
the longitudinal short-distance constraint where all virtualities are large
turns out to be matched only at the level of 81\%.
In this Letter we show that tensor mesons, whose contribution
to the muon $g-2$ has recently been found to be underestimated,
can fill this gap, because in holographic QCD
their infinite tower of excited mesons
only contributes to the symmetric longitudinal short-distance constraint.
Numerically, they give rise to a sizeable positive contribution from
the low-energy region below 1.5 GeV, a small one from
the mixed region, and a negligible one from
the high-energy region, which could explain the remaining gap between the
most recent dispersive and lattice results for the complete
hadronic light-by-light contribution.
\end{abstract}

\maketitle

\section{Introduction}

In its upcoming final result for the anomalous magnetic moment of the muon 
$a_\mu=(g-2)_\mu/2$, the Fermilab experiment \cite{Muong-2:2023cdq,Muong-2:2024hpx} 
is expected to reduce the current experimental error from
currently $22\times 10^{-11}$ to about half of this size, challenging
the Standard Model prediction \cite{Aoyama:2020ynm} which
needs further scrutiny and improvements regarding
its main source of uncertainties to match this precision.
At present, the main uncertainty arises from discrepancies
in the contribution from hadronic vacuum polarization 
\cite{Colangelo:2022jxc}, for which new lattice QCD results
\cite{Borsanyi:2020mff,Ce:2022kxy,ExtendedTwistedMass:2022jpw,FermilabLatticeHPQCD:2023jof,RBC:2023pvn,Boccaletti:2024guq,RBC:2024fic,Djukanovic:2024cmq} hold the promise of a more precise evaluation
while also removing most of the deviation between
the experimental result and the Standard Model value of the
Muon $g-2$ Theory Initiative
from 2020 \cite{Aoyama:2020ynm}. The latter involves
as second most important source of uncertainty the
effects of hadronic light-by-light scattering (HLbL),
at the time determined as 
$a_\mu^\text{HLbL} = 92(19) \times 10^{-11}$
\cite{Aoyama:2020ynm,Melnikov:2003xd,Masjuan:2017tvw,Colangelo:2017qdm,Colangelo:2017fiz,Hoferichter:2018dmo,Hoferichter:2018kwz,Gerardin:2019vio,Bijnens:2019ghy,Colangelo:2019lpu,Colangelo:2019uex,Pauk:2014rta,Danilkin:2016hnh,Jegerlehner:2017gek,Knecht:2018sci,Eichmann:2019bqf},
where the error needs to be reduced substantially as well
in view of the expected experimental improvements for $a_\mu$.

A new data-driven dispersive analysis \cite{Hoferichter:2024bae,Hoferichter:2024vbu}
has achieved a significant reduction of previous
uncertainties related to the contribution of axial vector
mesons and short-distance constraints (SDCs),
agreeing quite well with results obtained
with holographic models of QCD that have
indicated larger contributions than previously
assumed \cite{Leutgeb:2019gbz,Cappiello:2019hwh,Leutgeb:2021mpu,Leutgeb:2021bpo,Leutgeb:2022lqw,Colangelo:2024xfh,Leutgeb:2024rfs}, as also found
with resonance models \cite{Masjuan:2020jsf} and Dyson-Schwinger approximations
\cite{Eichmann:2024glq}.

Previous results \cite{Danilkin:2016hnh} for the contribution
of tensor mesons amounted to the almost negligibly small value of
$+0.64(13)\times 10^{-11}$
for the ground state tensor multiplet with a quark model ansatz
for the tensor transition form factors (TFFs), but
Ref.~\cite{Hoferichter:2024bae,Hoferichter:2024vbu} found that this needs to
be revised to a four times larger and negative value when a
formalism free of kinematical singularities introduced
recently in \cite{Hoferichter:2024fsj} is employed.
However, as a full dispersive treatment of tensor meson
contributions is not yet feasible, the same simple quark model,
though with different scale parameter, has been employed.

In this Letter we discuss tensor meson contributions to HLbL
in holographic QCD (hQCD) models whose predictions for the axial-vector
contributions have been validated numerically to a remarkable degree
by Ref.~\cite{Hoferichter:2024bae,Hoferichter:2024vbu}, and
we show that the infinite tower of tensor mesons plays an
important role in the longitudinal short-distance constraints (LSDC)
together with axial vector mesons. In the holographic approach,
the infinite towers of the latter
account fully for the Melnikov-Vainshtein SDC \cite{Melnikov:2003xd} on the longitudinal
amplitude, but universally contribute only 81\% of the symmetric LSDC where
all photon virtualities are large.

Fitting the tensor coupling such that the symmetric LSDC
is saturated in combination with the axial vector mesons
leads to a two-photon coupling of the ground-state tensor modes
in remarkable agreement with available data from the
BELLE experiment \cite{Belle:2015oin} 
for the singly-virtual TFF.
However, in contrast to the simple quark model employed in
Ref.~\cite{Hoferichter:2024bae,Hoferichter:2024vbu},
it predicts a positive contribution to $a_\mu$.
Moreover, the combined contribution of the
full tensor tower leads to a surprisingly large result,
$a_\mu^T\sim +11\times 10^{-11}$,
which could explain the remaining difference between
the new
dispersive result of \cite{Hoferichter:2024bae,Hoferichter:2024vbu}
and current lattice results for the
complete HLbL contribution of Refs.\ \cite{Chao:2021tvp,Chao:2022xzg, Blum:2023vlm,Fodor:2024jyn}.

\section{HLbL tensor and $(g-2)_\mu$ }

The central object in the calculation of the HLbL contribution to
$(g-2)_\mu$ is the HLbL tensor, i.e. the hadronic four-point function of electromagnetic currents
($q_4=q_1+q_2+q_3$)
\begin{align}
	\Pi^{\mu\nu\lambda\sigma}(q_1,q_2,q_3)&= -i \int d^4x \, d^4y \, d^4z \, e^{-i(q_1 \cdot x + q_2 \cdot y + q_3 \cdot z)} \notag\\
	&\hspace{-20pt}\times\langle 0 | T \{ j_\text{em}^\mu(x) j_\text{em}^\nu(y) j_\text{em}^\lambda(z) j_\text{em}^\sigma(0) \} | 0 \rangle
\end{align}
where
\begin{align}
	j_\text{em}^\mu = \bar q \Q \gamma^\mu q, \quad \Q = \text{diag}\left(\frac{2}{3}, -\frac{1}{3}, -\frac{1}{3}\right),
\end{align}
with 
$N_f=3$ quark fields $q = (u , d, s)^T$. 

A very useful 
representation of the HLbL tensor can be obtained using the 
Bardeen, Tung~\cite{Bardeen:1969aw}, and Tarrach~\cite{Tarrach:1975tu}
(BTT) decomposition
\beq\label{BTT}
\Pi^{\mu\nu\lambda\sigma} = \sum_{i=1}^{54} T_i^{\mu\nu\lambda\sigma} \Pi_i,
\eeq
with scalar functions $\Pi_i$ depending on the
virtualities $q_i^2$, and Lorentz structures 
$T_i^{\mu\nu\lambda\sigma}$ defined in \cite{Colangelo:2015ama}. This
decomposition is manifestly gauge invariant
and satisfies crossing symmetries. 

The contribution to $(g-2)_\mu$ is then given by
 the master equation~\cite{Colangelo:2017fiz} 
\begin{align}\label{master}
	a_\mu^\mathrm{HLbL}& = \frac{2 \alpha^3}{3 \pi^2} \int_0^\infty dQ_1 \int_0^\infty dQ_2 \int_{-1}^1 d\tau \sqrt{1-\tau^2} Q_1^3 Q_2^3\nonumber\\
	&\qquad\times\sum_{i=1}^{12}  T_i(Q_1,Q_2,\tau) \bar \Pi_i(Q_1,Q_2,Q_3)\,,
\end{align}
where twelve hadronic scalar functions $\bar \Pi_i$ built from
the $\Pi_i$ in \eqref{BTT} are  evaluated for the reduced kinematics
\begin{align}
(q_1^2,q_2^2,q_3^2,q_4^2)=(-Q_1^2,-Q_2^2,- Q_1^2 - 2 Q_1 Q_2 \tau - Q_2^2,0)\,.
\end{align}

Isolating single resonance exchange contributions to the $\bar \Pi_i$
in terms of their TFFs is a nontrivial task, which has recently been
extended to spin-1 and spin-2 resonances using a new optimized basis
\cite{Hoferichter:2024fsj}. In the case of tensor meson exchanges,
the absence of kinematic singularities is however given only
when the tensor TFFs have certain simplifications.
(Nevertheless, in triangle kinematics, the problem this poses
for the dispersive approach can be avoided in the strategy developed in \cite{Ludtke:2023hvz,Ludtke:2024ase}.)

A particular problem for hadronic models is the fact that TFFs of 
meson resonances have a power-law decay at large virtualities \HS\
that makes it impossible to satisfy the SDCs 
for the HLbL tensor following from the OPE
of pQCD by a finite number of resonances.
Especially important is the SDC derived for the longitudinal amplitude $\bar\Pi_1$
by Melnikov and Vainshtein \cite{Melnikov:2003xd} 
from the triangle anomaly in combination with the OPE
when two virtualities are much larger than the third,
\begin{align}
    \C_\mathrm{MV}=\lim_{Q_3\to\infty}\lim_{Q\to\infty}
    Q^2 Q_3^2 \bar\Pi_1(Q,Q,Q_3)= -\frac2{3\pi^2},
\end{align}
while the symmetric limit gives
\begin{align}
    \C_\mathrm{sym}=\lim_{Q\to\infty}
    Q^4 \bar\Pi_1(Q,Q,Q)=-\frac4{9\pi^2},
\end{align}
as proved (and extended) in the 
subsequent analyses of Refs.~\cite{Colangelo:2019lpu,Bijnens:2020xnl,Bijnens:2021jqo}.

\section{Holographic QCD}

Hadronic models in hQCD \cite{Sakai:2004cn,Erlich:2005qh,DaRold:2005mxj, Hirn:2005nr, Karch:2006pv} are inspired by the  original  AdS/CFT  duality between a  four-dimensional (4D) conformal large-$N_c$ gauge theory at strong coupling and a classical five-dimensional  (5D) field theory in a curved gravitational background with anti-de-Sitter metric \cite{Maldacena:1997re,Gubser:1998bc,Witten:1998qj}: for every quantum
operator ${\cal O}(x)$ of the 4D (strongly coupled) gauge theory, there exists a corresponding 5D  field $\phi(x,z)$ living in an (asymptotically AdS space, whose value on the conformal boundary (taken at $z=\epsilon\to0$) $\phi(x,0)\equiv\phi_0(x)$, is identified (modulo some specific powers of $\epsilon$), with the
four-dimensional source of ${\cal O}(x)$. In this way correlation functions of the 4D operators can be obtained by functional derivatives of the 5D Lagrangian with respect to the values of the 5D fields on the conformal boundary.

Correlation functions  of (conserved) $SU(3)_L\times SU(3)_R$ (or
$U(3)_L\times U(3)_R$) flavor chiral currents of QCD, in the large-$N_c$ limit,
 are obtained from 
a 5D  Yang-Mills action with coupling constant $g_5$
plus a Chern-Simons term
in a curved 5D AdS$_5$ space. 
In the simplest case that we shall consider here, in so-called hard-wall (HW) models, the extra fifth (holographic) dimension 
extends over the finite interval $z\in(0,z_0]$,
\begin{align}\label{eq:SYM5D}
S_{\rm YM} = -\frac{1}{4g_5^2} \int d^4x &\int_0^{z_0} dz\,
\sqrt{-g}\, g^{PR}g^{QS}\notag\\
&\times\text{tr}\left(\mathcal{F}^\mathrm{L}_{PQ}\mathcal{F}^\mathrm{L}_{RS}
+\mathcal{F}^\mathrm{R}_{PQ}\mathcal{F}^\mathrm{R}_{RS}\right),
\end{align}
where $P,Q,R,S=0,\dots,3,z$ and $\mathcal{F}_{MN}=\partial_M \mathcal{B}_N-\partial_N \mathcal{B}_M-i[\mathcal{B}_M,\mathcal{B}_N]$, $\mathcal{B}_N=L_N,\,R_N$ being 5D gauge fields  transforming under $U(3)_{L, R}$ respectively; vector and axial-vector fields are given by ${V}_{\mu}=\frac{1}{2}( {L}_{\mu} + {R}_{\mu})$ and $ {A}_{\mu}=\frac{1}{2}( {L}_{\mu} - {R}_{\mu})$. 

Interactions with photons involve the vector bulk-to-boundary propagator $\J$, which is a solution of the vector 5D equation of motion for generic value of the 4D momentum  $q^2$ satisfying the  boundary conditions 
$\J(q,0)=1$ and $\partial_z \J(q,z_0)=0$. 
For Euclidean momentum $q^2=-Q^2$, it is given in HW models by 
\be\label{HWVF}
\J(Q,z)=
Qz \left[ K_1(Qz)+\frac{K_0(Q z_0)}{I_0(Q z_0)}I_1(Q z) \right],
\ee
where $I$ and $K$ denote modified Bessel functions.

According to the holographic recipe, 
the 4D  vector current two-point function follows from the
asymptotic behavior of (\ref{HWVF}), and matching it to
the leading-order pQCD result for $Q^2\to\infty$
\cite{Erlich:2005qh},
\be\label{PiVas}
\Pi_V(Q^2)=-\frac1{g_5^2 Q^2} \left( \frac1z \partial_z \J(Q,z) \right)\Big|_{z\to0} \sim -\frac{N_c}{24\pi^2}\ln Q^2, 
\ee
determines $g_5^2=12\pi^2/N_c=(2\pi)^2$ for $N_c=3$. 

Single resonances appear as the coefficients of the expansion of the 5D fields in terms of a series of normalizable eigenfunctions. In the case of the vector field the masses are given by $m_n=\gamma_{0,n}/z_0$, where $\gamma_{0,n}$ is the $n$th zero of the Bessel function $J_0(\gamma_{0,n})=0$. 
Identifying the mass of the lowest vector resonance with the mass of the $\rho$ meson $m_1=\gamma_{0,1}/z_0=M_\rho=775.26$ MeV fixes
$z_0=\gamma_{0,1}/m_1=3.102 \, \text{GeV}^{-1}$.

A rich dynamics is contained in the sector of axial gauge fields, 
which yields a TFF for axial vector mesons \cite{Leutgeb:2019gbz}
in agreement
with the Landau-Yang theorem \cite{Landau:1948kw,Yang:1950rg} and with the (subsequently derived) doubly virtual
asymptotic behavior deduced from QCD light-cone expansions
\HS.
Summing over the infinite tower
of axial vector mesons one finds that the MV-SDC is
realized in the form \cite{Leutgeb:2019gbz,Cappiello:2019hwh,Leutgeb:2021mpu}
\begin{align}
    \C_\mathrm{MV}^\mathrm{A}=-\left(\frac{g_5}{2\pi}\right)^2\frac1{\pi^2}\int_0^\infty d\xi\, \xi [\xi K_1(\xi)]^2=-\frac2{3\pi^2}\left(\frac{g_5}{2\pi}\right)^2.
\end{align}
The symmetric LSDC, on the other hand, is reproduced as far as the power-law behavior
is concerned, but the coefficient therein is 19\% too small:
\begin{align}\label{Pi1AVsymlimit}
\C_\mathrm{sym}^\mathrm{A}&=-\left(\frac{g_5}{2\pi}\right)^2\frac1{\pi^2}
\int_0^\infty d\xi  
\,\xi [\xi K_1(\xi)]^3 \nonumber\\
&=-0.8122\times\frac{4}{9\pi^2}\left(\frac{g_5}{2\pi}\right)^2.
\end{align}

\section{Tensor mesons in hQCD}

Following \cite{Katz:2005ir,Mamedov:2023sns,Colangelo:2024xfh}, the tensor meson is introduced in the model as a deformation of the 4D part of the AdS metric
\begin{equation}\label{eq:metricfluctuation}
    ds^2=g_{MN}dx^M dx^N =\frac{1}{z^2} (\eta_{\mu\nu}+h_{\mu\nu}) dx^\mu dx^\nu-\frac{1}{z^2} dz^2\,,
\end{equation}
whose dynamics is governed by the 5D Einstein-Hilbert action
\be
S_\mathrm{EH} = -2k_T\int d^5x \,\sqrt{g} \, (\mathcal{R}+2\Lambda)
\ee
with $k_T$ determining the strength of the coupling to other fields.

Tensor mesons arise as
solutions for traceless-transverse metric fluctuation 
$h_{\mu\nu}=\epsilon_{\mu\nu}^T(q)h_n(z)$
with $h_n(0)=0=h_n'(z_0)$. 
Requiring canonical normalization ${k_T} \int dz z^{-3} (\partial_z h_n)^2={(m^T_n)^2}$, they are given explicitly by
\beq
    h_n(z) = \displaystyle\frac{\sqrt{2/k_T}}{{z_0}
   J_2\left(\gamma_{1,n}\right)} z^2 J_2(\gamma_{1,n}z/z_0),
\eeq
yielding an infinite tower of (flavor singlet) tensor meson modes
with masses fixed by the zeros of the Bessel function $J_1$,
\be
m_n^T/M_\rho=\gamma_{1,n}/\gamma_{0,1}=
1.593,\, 2.917,\, 4.230,\, \ldots 
\ee
The lowest tensor resonance has thus a mass $m^T_1=1.235$ GeV, which
is only $3\%$ below the physical value $M_{f_2}=1.2754$ GeV of $f_2(1270)$.%
\footnote{To distinguish physical and model parameters, we use upper-case
$M$ for experimental values of masses and lower-case $m$ for corresponding
mass parameters given by the hQCD model.}


\begin{figure}[t]
\includegraphics[width=0.45\textwidth]{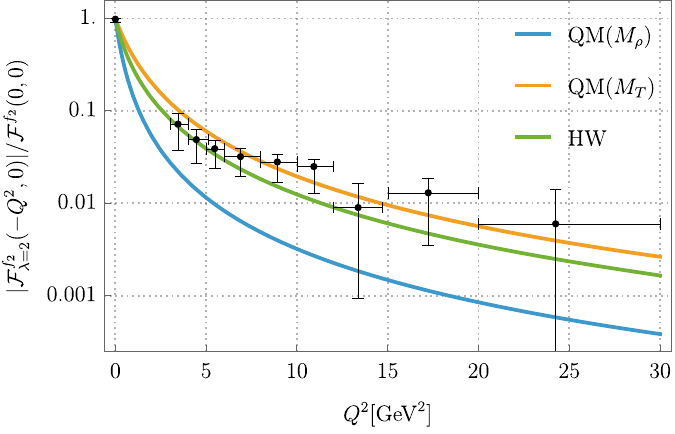}
\caption{Comparison of singly virtual tensor TFFs for helicity
$\lambda=2$ with Belle data \cite{Belle:2015oin} for the $f_2(1270)$, normalized by
$\F^{f_2}(0,0)\equiv\sqrt{\F^T_1(0,0)^2+\F^T_2(0,0)^2/24}=\sqrt{5 \Gamma_{\gamma\gamma}/(\pi\alpha^2 M_T)}$.
HW is the hard-wall hQCD result, QM($\Lambda_T$) refers to the
quark model of Ref.~\cite{Schuler:1997yw}, with scale parameter $\Lambda_T$ set to either $M_\rho$ \cite{Hoferichter:2024bae} or $M_T=M_{f_2}$ \cite{Hoferichter:2020lap,Schuler:1997yw}.
(Figure taken from \cite{Cappiello:2025fyf}, which also shows
the comparison for the remaining helicities.)
\label{fig:belle}}
\end{figure}

A single tensor mode gives rise to the $T\to\gamma^*\gamma^*$
amplitude
\begin{align}
    \M^{\mu \nu \alpha \beta}= T_1^{\mu \nu \alpha \beta} \frac{1}{m_T}\mathcal{F}_1+ T_3^{\mu \nu \alpha \beta}\frac{1}{m_T^3} \mathcal{F}_3
\end{align}
with structure functions (following the notation of \HS)
\begin{align}
    &\mathcal{F}^T_1(-Q_1^2,-Q_2^2)/m_T\nonumber\\
    &=- \frac{1}{g_5^2} \text{tr}\Q^2\int \frac{dz}{z} h_n(z) \J(z,Q_1)\J(z,Q_2), \\
       &\mathcal{F}^T_3(-Q_1^2,-Q_2^2)/m_T^3\nonumber\\
       &=-\frac{1}{g_5^2} \text{tr}\Q^2\int \frac{dz}{z} h_n(z) \frac{\partial_z\J(z,Q_1)}{Q_1^2} \frac{\partial_z\J(z,Q_2)}{Q_2^2}.
\end{align}

Evaluating this for the lowest tensor mode in the singly
virtual case where experimental data are available \cite{Belle:2015oin}
for the $f_2(1270)$ tensor meson,
one obtains a remarkably good agreement as shown in
Fig.\ \ref{fig:belle}, where it is compared with
the quark model (QM) ansatz \cite{Schuler:1997yw} with
dipole scale $\Lambda_T$ set to either $M_\rho$ as in \cite{Hoferichter:2024bae} or $M_T$ as in \cite{Hoferichter:2020lap,Schuler:1997yw}.

Summing over the infinite tower of tensor modes corresponds to
a Witten diagram with bulk-to-bulk tensor propagator involving
\be
G(z,z';q^2)=  \sum_0^{\infty} \frac{h_n(z)h_n(z')}{q^2-(m^T_n)^2}
\ee
which at $q^2=0$ has the simple form
\be
    G(z,z';0)=-\frac{1}{4}\text{min}(z^4,z'^4).
\ee
For the longitudinal amplitude this yields
\begin{align}\label{Pi1QQQ3}
      \bar{\Pi}_1&(Q,Q,Q_3)= -\frac{4}{k_T} (\frac{\text{tr}\Q^2}{g_5^2})^2 \int_0^{z_0} \frac{dz dz'}{z z'}\nonumber\\
      & \times\J(z,Q) \J(z',Q) \frac{ \partial_{z'}\J(z',Q_3)}{Q_3^2} \partial_{z'}G(z,z';0).
\end{align}
In \cite{Cappiello:2025fyf}
we have given the 
results for the individual tensor contributions to $\bar\Pi_i$,
evaluated alternatively with only the pole contribution as defined by the
dispersive approach \cite{Hoferichter:2024fsj} and without this restriction,
which differ strongly.
Summing over the
whole tower of tensor modes 
leads to exactly the same result \eqref{Pi1QQQ3},
whereas
in the case of axial vector mesons there is equivalence
between the two possibilities already at the
level of individual modes.

\begin{figure}[t]
\bigskip
\includegraphics[width=0.45\textwidth]{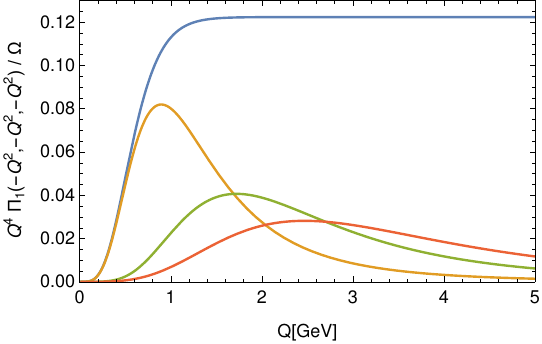}
\caption{Contribution of tensor mesons to the symmetric longitudinal
SDC: $Q^4 \,\hat\Pi_1(-Q^2,-Q^2,-Q^2)$ with full tensor bulk-to-bulk propagator (blue)
and the contributions from the first three modes (orange, green, red),
normalized to the full symmetric LSDC value $\Omega=-4/(9\pi^2)$
with $k_T$ fixed by \eqref{kTpQCD}.
By contrast, adding tensor contributions to Fig.~4 in \cite{Leutgeb:2019gbz}, which
shows how the tower of axial vector mesons saturates the MV-SDC in the
asymmetric limit $Q_3\ll Q_1\sim Q_2 \to\infty$, those contributions
would not be discernible as different from zero (cf.~Ref.~\cite{Cappiello:2025fyf}.)
\label{fig:Pi1SymSDC}}
\end{figure}

In the limit $Q_3\ll Q\to\infty$ the result \eqref{Pi1QQQ3} 
decays faster than $1/Q^2$, giving
no contribution to the MV-SDC.
But in the symmetric limit $Q_3=Q\to\infty$ one obtains
(cf.\ Fig.\ \ref{fig:Pi1SymSDC})

\begin{align}
    \bar{\Pi}_1(Q,Q,Q)\to&
    \frac{4}{Q^4}\frac{1}{k_T} (\frac{\text{tr}\Q^2}{g_5^2})^2\int_0^\infty d\xi K_1(\xi) 
    \int_0^\xi d\xi' \xi'^3 \nonumber\\
    &\qquad\qquad\qquad\times K_1(\xi')\, \partial_{\xi'}\left[\xi' K_1(\xi')\right],
    \nonumber\\
&=\frac{4}{Q^4}\frac{1}{k_T} (\frac{\text{tr}\Q^2}{g_5^2})^2 \times (-0.15285). 
\end{align}
This is $12.23 \%$ of the OPE result for $N_c=N_f=3$
when $k_T$ is matched through the energy-momentum tensor two-point function
as done originally in \cite{Katz:2005ir}, i.e.,
\be\label{kTpQCD}
k_T\equiv\frac1{g_f^2}=\left(\frac{N_c N_f}{80\pi^2}+\frac{N_c^2-1}{40\pi^2}\right)
=\frac{5}{16\pi^2}.
\ee
Together with the contribution from the axial vector mesons,
the symmetric SDC is matched at the level of $81.22+12.23= 93.45 \%$,
and this could be increased to 98.2\% if \eqref{kTpQCD}
is reduced to the term proportional to $N_c^2$
that would be selected in a large-$N_c$ expansion.

However, this nearly perfect match is somewhat coincidental, since
it does not reproduce the correct large-$N_c$ scaling.
The OPE result is proportional to $N_c \text{tr}\Q^4$, whereas
the above result involves $N_c^0 (\text{tr}\Q^2)^2$ because
$g_5^2=12\pi^2/N_c$.
The correct behavior could be obtained by a multiplet of
tensor contributions with $k_T\propto N_c$ as provided by the quark part
of the energy-momentum tensor, contributing proportional to the sum
$\sum_{a=0,3,8}(\text{tr}\Q^2 T^a)^2=\frac12 \text{tr}\Q^4$.
Instead of fixing $k_T$ by \eqref{kTpQCD}, one could
then determine it by matching the combined contribution of
axial vector and tensor meson contributions to the symmetric LSDC.
With degenerate tensor mesons as given by the AdS/QCD model,
the result is simply an extra overall factor compared to the above result.
For $g_5^2=(2\pi)^2$ as given by the OPE of the vector correlator,
the tensor contribution to the HLbL amplitude and thus to $a_\mu$
needed to be enhanced by a factor 1.536; if one chooses a reduced
$g_5^2=0.894(2\pi)^2$ by fitting $F_\rho$ as proposed in \cite{Leutgeb:2022lqw},
the contribution from tensor mesons is enhanced by a factor $1/0.894^2$ relative
to the axial vector contributions so that another factor of 1.097 suffices
to have the correct ratio between symmetric and MV SDC.
In these two scenarios, the contributions from the tensor mesons to $a_\mu$
would have to be enhanced simply by a factor 1.536 and 1.373, respectively,
compared to the initial choice \eqref{kTpQCD} together with $g_5^2=(2\pi)^2$.
(This also mitigates somewhat
the numerical deficit that was obtained in \cite{Katz:2005ir} for the ratio
$\Gamma(f_2\to\pi\pi)/\Gamma(f_2\to\gamma\gamma)$ \cite{Cappiello:2025fyf}.)

In fact, matching the lowest tensor meson pole contribution 
in the AdS/QCD model with \eqref{kTpQCD}
to mass
and two-photon decay rates of the experimentally observed 
$f_2(1270), a_2(1320), f_2'(1525)$
one obtains 
also a result that is about a factor 1.5 higher.  
The two options for fitting $g_5^2$, OPE-fit and $F_\rho$-fit, 
and then fixing $k_T$
through the OPE ratio of the symmetric and asymmetric LSDC
give a result that tightly brackets
the one obtained by matching the experimental data
for $f_2, a_2, f_2'$, see Table \ref{tab:T1}.
Here more than 90\% of the tensor contributions to $a_\mu$
are from the
low-energy region $Q_i\le Q_0=1.5$ GeV, while contributions
from all $Q_i>Q_0$ are below $10^{-12}$.
(For more details see \cite{Cappiello:2025fyf}.)

\begin{table}[t]
    \centering
    \begin{tabular}{c|c|c|c|c|c}
    \toprule
    $k_T$ & $M_T$ [GeV] & $\Gamma_{\gamma \gamma}$ [keV] &IR &Mixed & $a_\mu$ $[10^{-11}]$\\
    \colrule
    by $\Gamma_{\gamma\gamma}$ & 1.2754(8) & 2.65(45) & $2.28$ & $0.16$ &  2.4(4) \\
    & 1.3182(6) & 1.01(9) & $0.85$ & $0.05$ &  0.9(1) \\
    & 1.5173(24) & 0.08(2)   & 0.06 & $0.003$ &  0.06(2)\\
    & $f_2+a_2+f_2'$ & & $3.19$ & $0.21$ &  3.4(4) \\
    \hline
    $F_\rho$ fit & 1.235  & 2.3+0.8+0.2 & 2.93 & 0.23 & 3.17 \\
    OPE fit & 1.235 & 2.6+0.9+0.2 & 3.28 & 0.25 & 3.55 \\
    \botrule
    \end{tabular}
    \caption{$a_\mu$ results for the tensor meson pole contributions obtained by inserting the hQCD results for $\F^T_{1,3}$ in the formulae obtained in \cite{Hoferichter:2024fsj} within the dispersive approach, 
    when experimental values of masses and two-photon decay rates
    are matched (upper four lines), compared to the results
    in the AdS/QCD model when $k_T$ is fixed by the symmetric SDC
    with $g_5$ alternatively determined by a fit of $F_\rho$ or
    the OPE of the vector correlator. The columns IR and Mixed
    give the contributions from the low-energy (IR) region
    defined by $Q_i\le Q_0=1.5$ GeV and from the region where
    one or two virtualities are above $Q_0$. (Contributions from
    the region where $Q_i>Q_0$ for all $Q_i$ are below $10^{-13}$.)
    In the last two lines $\Gamma_{\gamma\gamma}$ for the (degenerate) AdS/QCD result with $m^T_1=1.235$ GeV has been split as (69+25+6)\% corresponding to
    ideal mixing for $f_2+a_2+f_2'$.}
    \label{tab:T1}
\end{table}

Summing over the infinite tower of excited tensor modes
by using the full bulk-to-bulk propagator 
still gives tiny (a few $10^{-12}$) contributions from the high-energy
region $Q_i>Q_0$, but rather large contributions
in the IR region
\bea
&&a_\mu^{T,\mathrm{total}}\times 10^{11} =\begin{cases}
    11.1 & \text{($F_\rho$-fit)} \\
    12.4 & \text{(OPE-fit)}
\end{cases},\nonumber
\\
&&a_\mu^{T,\mathrm{total}}|_\mathrm{Mixed}\times 10^{11} =\begin{cases}
     1.9 & \text{($F_\rho$-fit)} \\
     2.1 & \text{(OPE-fit)}
\end{cases},\nonumber
\\
&&a_\mu^{T,\mathrm{total}}|_\mathrm{IR}\times 10^{11}=\begin{cases}
    8.5 & \text{($F_\rho$-fit)} \\
    9.5 & \text{(OPE-fit)}
    \end{cases}.
\eea
In line with the ``best-guess'' model for the hQCD axial-sector
contributions of \cite{Leutgeb:2022lqw} we would adopt
the $F_\rho$-fit case as central value, the OPE-fit case
as upper, and the original parameter choice of \cite{Katz:2005ir}
as lower values for a theoretical error estimate within hQCD,
\bea
&&a_\mu^{T,\mathrm{total}}=11.1^{+1.3}_{-3.0}\times 10^{-11},\nonumber\\
&&a_\mu^{T,\mathrm{total}}|_\mathrm{IR}=8.5^{+1.0}_{-2.3}\times 10^{-11}.
\eea

A contribution in the IR region
of this size from tensor mesons 
would move the central value
for $a_\mu^\mathrm{HLbL,total}$ obtained therein 
in a dispersive approach from\footnote{Ref.\ \cite{Hoferichter:2024bae,Hoferichter:2024vbu}
obtained $-2.5\times10^{-11}$ for the pole contribution of the
ground-state tensor multiplet in the IR region, 
which in the hQCD case with $F_\rho$-fit reads $+2.9\times10^{-11}$ (see
Table \ref{tab:T1}). Ref.\ \cite{Hoferichter:2024bae,Hoferichter:2024vbu}
also includes additional effective poles to account for
other higher intermediate states. The amount of their contribution
($+2.0\times10^{-11}$)
is comparable to the contributions of excited pseudoscalar and
axial vector states in the hQCD model of \cite{Leutgeb:2022lqw} ($+2.3\times10^{-11}$
for $F_\rho$-fit). The hQCD ($F_\rho$-fit) result for the contribution of excited tensors to the
IR region amounts to an extra $8.5-2.9=5.6\times10^{-11}$ in this category.}
101.9(7.9) to about $113\times 10^{-11}$.
This would then agree perfectly with the recent lattice results
of $109.6(15.9)$, $124.7(14.9)$, and $125.5(11.6)\times10^{-11}$ of
the Mainz group \cite{Chao:2021tvp,Chao:2022xzg}, the RBC-UKQCD group \cite{Blum:2023vlm}, and the BMWc lattice group \cite{Fodor:2024jyn}, respectively.

As shown in \cite{Cappiello:2025fyf}, the sizable \textit{positive} contribution
from tensor mesons is a consequence of 
the presence of the structure function $\F_3^T$, which
is not constrained by existing data on singly virtual TFFs for tensor mesons.
Dropping it would give even larger but
negative results for the tensor contributions,
but within the holographic approach this cannot be done
without violating the symmetries of AdS; moreover,
because of chiral symmetry, dropping the corresponding term in the vector part of the action would require
one to drop also the same kind of term for axial vectors, completely losing the coupling between tensor mesons
and pions in the Hirn-Sanz \cite{Hirn:2005nr} model.

In \HS, 
employing a light-cone expansion generalizing the Brodsky-Lepage approach  \cite{Lepage:1979zb,Brodsky:1981rp},
it was found that
$\F_{2,4,5}^T$ are asymptotically of comparable
magnitude as $\F_3^T$, while they do not arise in the present hQCD model. Moreover, the
asymptotic asymmetry functions obtained for $\F_{1,3}^T$ differ \cite{Cappiello:2025fyf}, while
there is agreement between the holographic results and the results of \HS\ in the case of axial-vector mesons \cite{Leutgeb:2019gbz}.
This calls for further studies, in particular regarding the remaining SDCs \cite{Bijnens:2020xnl,Bijnens:2021jqo}
on the HLbL amplitude.

However, 
the significant contributions obtained for the tensor-meson
contributions to $a_\mu^\mathrm{HLbL}$ in hQCD arise primarily from the
low-energy region. As shown in Appendix A of \cite{Cappiello:2025fyf},
the two structure functions $\F_{1,3}^T$ are directly related to
the two structures of kinetic and mass terms of vector meson fields
interacting with tensor fields, making it physically plausible that
both contribute importantly
at low energies.
In the case of $\F_3^T$, this prediction of hQCD
can be tested in the doubly virtual $T\to\gamma^*\gamma^*$
processes, for which experimental data 
will hopefully become available in the future.
Already now it is clear from our hQCD results that there is
room for further sizeable contributions in the dispersive approach of \cite{Hoferichter:2024bae,Hoferichter:2024vbu} coming from the tensor meson sector,
which will require additional experimental input for a conclusive treatment.

\begin{acknowledgments}
We would like to thank Martin Hoferichter for very useful discussions. 
This work has been supported by the Austrian Science Fund FWF, Grant-DOI
10.55776/PAT7221623. L.C. acknowledges the support of the INFN research project
ENP (Exploring New Physics).
\end{acknowledgments}

\bibliographystyle{JHEP}
\bibliography{SM}

\end{document}